\documentclass[twocolumn,amsmath,amssymb]{revtex4}

\usepackage{graphicx}
\usepackage{dcolumn}
\usepackage{bm}

\graphicspath{{Figures/}}
\setkeys{Gin}{width=\linewidth}

\newcommand*{\fref}[1]{Fig.\,\ref{#1}}
\newcommand*{\Fref}[1]{Figure\,\ref{#1}}
\newcommand*{\rmd}{\mathrm{d}}
\newcommand*{\diff}[2]{\ensuremath{\frac{\rmd #1}{\rmd #2}}}

\newcommand*{\sub}[1]{\ensuremath{_{\rm #1}}}
\newcommand*{\ttop}[1]{^{\rm #1}}
\newcommand*{\Fts}{\ensuremath{F\sub{ts}}}
\newcommand*{\nHall}{\ensuremath{n\sub{Hall}}}
\newcommand*{\nSdH}{\ensuremath{n\sub{SdH}}}

\begin{document}

\title{Kelvin Probe Spectroscopy of a Two-Dimensional Electron
  Gas Below 300$\,\textrm{mK}$}

\author{T.\ Van\v cura, $^{a)}$ S.\ Ki\v cin, $^{a)}$ T.\ Ihn, $^{a)}$ K.\ 
  Ensslin,$^{a)}$ M.\ Bichler,$^{b)}$ and W.\ Wegscheider$^{c)}$}

\affiliation{$^{a)}$Laboratory of Solid State Physics, ETH Zurich,
  8093 Zurich, Switzerland,\\$^{b)}$Walter Schottky Institute,
  85748 Garching, Germany, \\$^{c)}$Applied and
  Experimental Physics, University of Regensburg, 93040 Regensburg,
  Germany}

\begin{abstract}

  A scanning force microscope with a base temperature below
  $300\,{\rm mK}$ is used for measuring the local electron density
  of a two-dimensional electron gas embedded in an Ga[Al]As
  heterostructure. At different separations between AFM tip and
  sample, a dc-voltage is applied between the tip and the electron
  gas while simultaneously recording the frequency shift of the oscillating tip. 
Using a plate capacitor model the local electron
  density can be extracted from the data. The result coincides
  within 10$\,$\% with the data obtained from transport
  measurements.

\end{abstract}

\maketitle


The electron density of two-dimensional electron gases (2DEGs) is
usually determined by magnetotransport experiments. The carrier
density can be either extracted from the low-field slope of the Hall
resistance or from the 1/B periodicity of Shubnikov-de Hass
oscillations.  Also C--V profilometry \cite{ambridge:1973, blood:1986,
  baxandall:1971, copeland:1969} and magnetocapacitance experiments
\cite{mosser:1986, smith:1986} are versatile tools to detect the
electron density in an electron gas located below a metallic top
gate electrode.

Here we set out to use a Kelvin probe technique in order to measure
the local electron density in a 2DEG below the conductive tip of an
atomic force microscope.

The two-dimensional electron gas investigated is embedded in a
Ga[Al]As heterostructure with the electrons buried $40\,{\rm nm}$
below the surface.  No mesa structure was imprinted. Ohmic contacts
at the sample edges allow to measure the 4-terminal resistances at
low temperatures and to determine the carrier density through
transport measurements.

The sample is mounted in a home built scanning probe microscope (SPM) situated in a
$^{3}$He-cryostat \cite{ihn:2002} where an operating temperature
below 300$\,$mK is reached routinely.  Scanning is performed with an
electrochemically etched metallic tip attached to the end face of
one prong of a piezoelectric quartz tuning fork \cite{guthner:1989,
  karrai:1995, edwards:1997, rychen:2000a, giessibl:2000b}.

Optical detection of the cantilever deflection is not suitable for
our purposes, because the sample's electronic properties are
sensitive to light (persistent photoeffect).  Therefore the setup
relies on a piezoelectric measurement of the tip oscillation
utilizing a phase-locked loop measuring the change in resonance
frequency upon changes in the tip-sample interaction
\cite{albrecht:1991,duerig:1992}. The relative accuracy of the frequency
detection is better than $10^{-7}$ \cite{rychen:2000a, rychen:2000b,
  ihn:2002}.

In a dynamic mode SPM at small tip oscillation amplitudes, the measured frequency shift $\Delta f$ does
not directly reflect the force \Fts\ acting on the cantilever, but
rather the force gradient \cite{albrecht:1991}
\begin{equation}
  \Delta f \propto \diff{\Fts}{z} =: \Fts^{\prime}\,.
  \label{eq:deltaf}
\end{equation}


By applying a dc-voltage between the metallic tip and the sample,
the density is modified. In a metallic system one expects a
parabolic voltage dependence of the force gradient. The curvature of
the parabola is determined by the capacitive coupling between tip
and sample. The position of the apex of the parabola determines the contact
potential difference $U\sub{CPD}$ of the two metals. This method is
generally known as Kelvin
probe \cite{nonnenmacher:1991,nabhan:1997,jacobs:1999}.

\begin{figure}
  \begin{center}
    \includegraphics{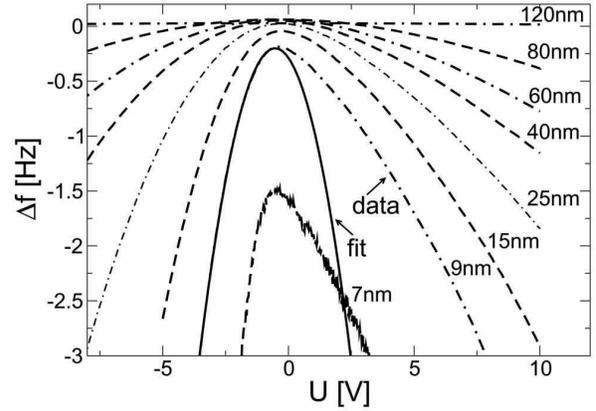}
    \caption{Kelvin probe at different tip-sample distances. 
      On the vertical axis the frequency shift $\Delta f$ is plotted
      versus the bias voltage $U$ on the horizontal axis.  
      The solid line indicates the parabolic fit to the data taken
      at $9{\,\rm nm}$ tip-sample separation. We choose the sign of
      the voltage always with respect to the 2DEG, i.e., a negative
      voltage indicates that the negative contact is connected to
      the electron gas.  }
    \label{fig:kelvin1}
  \end{center}
\end{figure}

\Fref{fig:kelvin1} shows Kelvin probe data measured at different
tip-sample distances with a 2DEG underneath the tip.  The curvature of the measured curves is different for positive
and negative voltages as indicated by the fit to the $9\,$nm curve.
The reason lies in the depletion of the electron gas underneath the
tip for positive voltages. This changes the tip-sample capacitance
and thus reduces the force coupling.


The electrostatic force gradient $\Fts^{\prime}$ between tip and
sample responds to a change in the bias voltage $U$ as
\begin{equation}
    \Fts^{\prime} = 
    \frac{1}{2}\diff{^{2}C(z, U)}{z^{2}}(U-U\sub{CPD})^2\,,
    \label{eq:elec-force2}
\end{equation}
where
$C(z,U)$ is the tip-sample
capacitance. For metallic samples $C(z,U)$ is independent of $U$ and the maximum of the parabola is shifted in voltage by
$U\sub{CPD}$.  The experiments were performed after a series of
image scans.
The tip
was not very sharp, increasing the tip-sample capacitance.

The solid line in \fref{fig:kelvin1} shows a parabolic fit to the
9$\,$nm trace. At negative bias voltages the fit was made to agree
well with the data, i.e. in the regime, where the electron gas is
not depleted. At positive voltages, however, where the electron gas
becomes depleted and the tip-sample coupling is weakened due to the
voltage dependence of $\mathrm{d}^{2}C(z,U)/\mathrm{d}z^{2}$, the
curvature is reduced. Already in \fref{fig:kelvin1}, a dependence
between the point, where the depletion sets in, and the tip-sample
distance can be seen. This point of depletion will be the focus of
the following discussion.

\begin{figure}
  \begin{center}
    \includegraphics{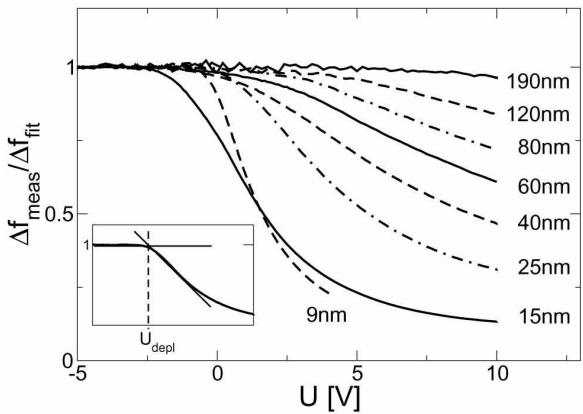}
    \caption{  \label{fig:relDf}
      Relative Kelvin probe.  The data from \fref{fig:kelvin1} are
      divided by their respective parabolic fit.  \textit{Inset:\/}
      The position where the depletion sets in, $U\sub{depl}$, is
      defined as indicated as the point where the tangent to the
      curve crosses the $y=1$ axis. }
  \end{center}
\end{figure}
\begin{figure*}[tbp]
  \centering \includegraphics[width=0.45\linewidth]{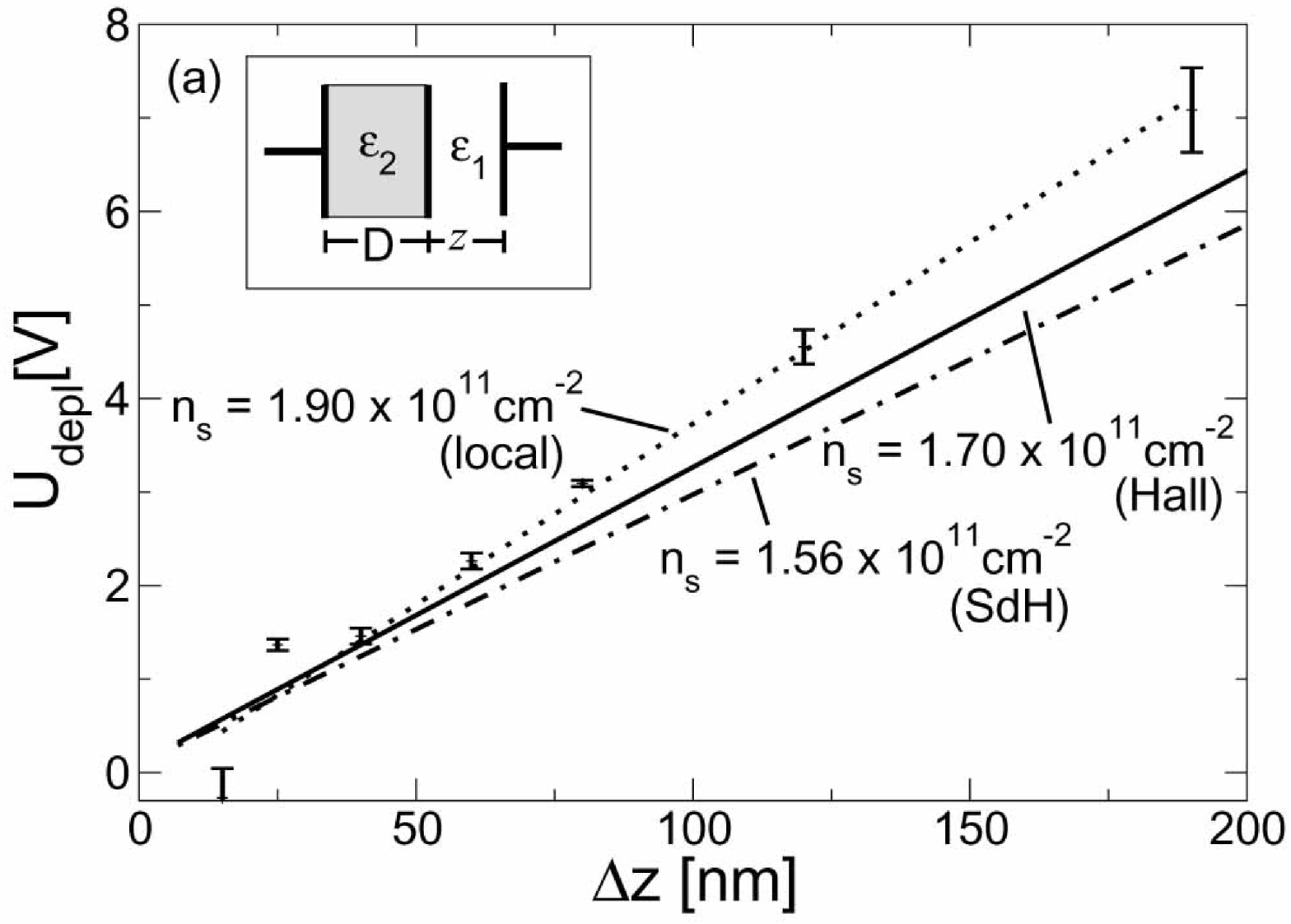}
  \hspace{0.01\linewidth}
  \includegraphics[width=0.45\linewidth]{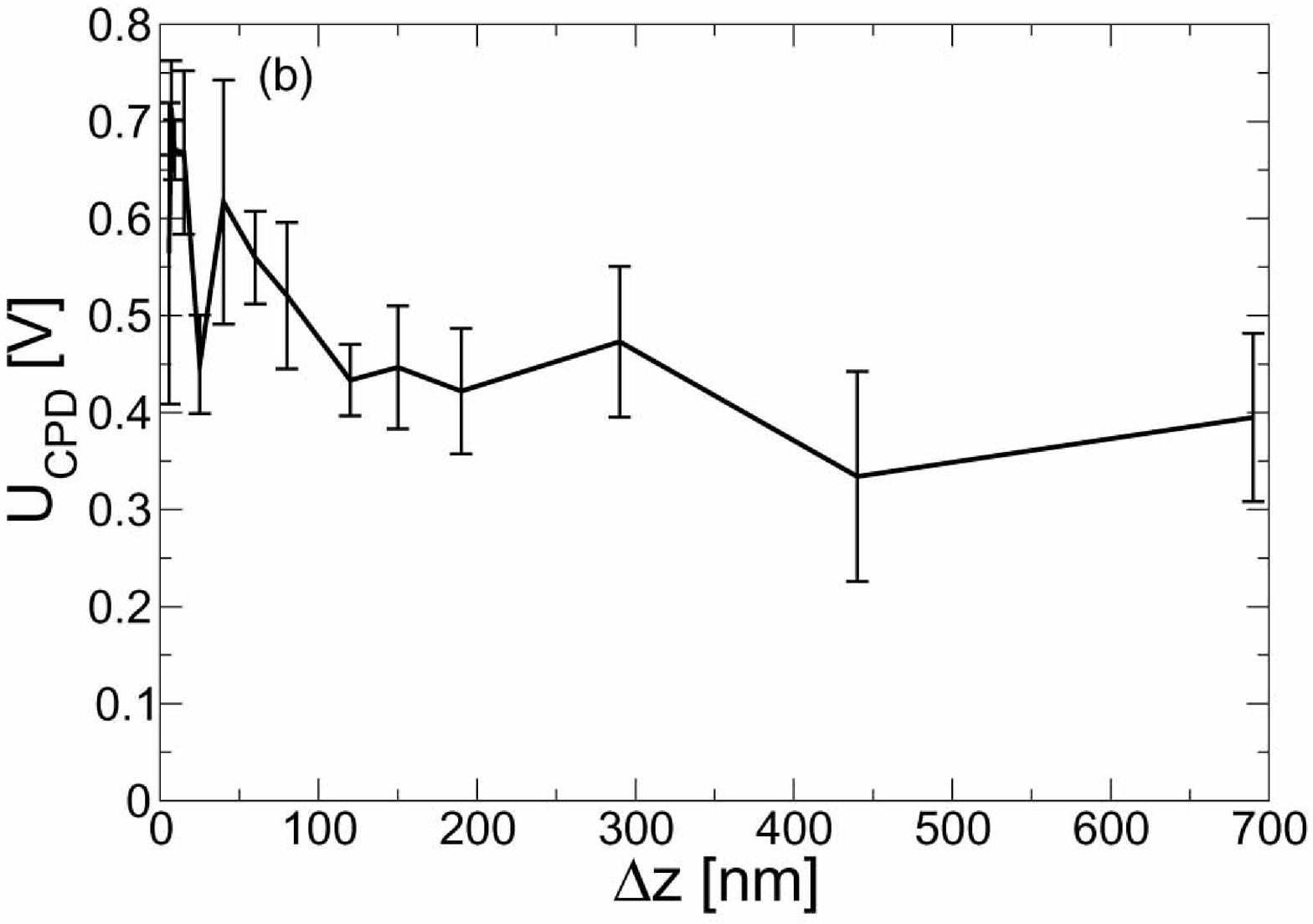}

  \caption{\label{fig:delta-mu}
    \textit{a)\/} Depletion voltage plotted versus the tip-sample
    distance at which the Kelvin probe was recorded. The electron
    density extracted from the transport data is $n_{\rm Hall}=
    1.70\cdot10^{11}\,{\rm cm}^{-2}$ and $n_{\rm SdH} =
    1.56\cdot10^{11}{\rm cm}^{-2}$.  This leads to the solid lines
    in the graph. Inset: Model geometry.
    \textit{b)} Contact potential difference $U\sub{CPD}$ plotted as
    a function of $\Delta z$.}
\end{figure*}

In order to determine $U\sub{depl}$ quantitatively the ratio $\Delta
f\sub{meas}/\Delta f\sub{fit}$ is evaluated as a function of $U$
(see \fref{fig:relDf}).  We define the depletion voltage
$U\sub{depl}$ as the position of the knee in $\Delta
f\sub{meas}/\Delta f\sub{fit}$ determined as shown in the inset
of \fref{fig:relDf}.  These depletion voltages are plotted versus
the respective tip-sample separation in \fref{fig:delta-mu}a. The
broken line is a linear fit to the data.

We approximate our setup with a plate capacitor model to
extract a local electron sheet density $n\sub{s}$ from this data.
One capacitor plate, the tip, resides at a distance $z$ above the
sample surface.  The two-dimensional electron gas is buried
underneath a GaAs cap layer of thickness $D$.  The dielectric
constants are $\epsilon_{1} = 1$ for the vacuum and $\epsilon_{2} =
12$ for GaAs, respectively (see inset in \fref{fig:delta-mu}a).

For $U<U\sub{depl}$ the total capacitance $C\sub{tot}(z)$ is assumed to be independent of voltage and given by
\begin{displaymath}
  \frac{C\sub{tot}(z)}{A} = \frac{\epsilon_{0}\epsilon_{1}\epsilon_{2}}
      {\epsilon_{2}z + \epsilon_{1}D}\,,
\end{displaymath}
where $A$ is the area of the plates. 

The charge density in the 2DEG is given by
\begin{displaymath}
  e\, n\sub{s}(U) = \frac{\Delta Q}{A} = -\frac{C\sub{tot}}{A}(U-U\sub{CPD})+e\,n\sub{s}^{(0)},
\end{displaymath}
where $n\sub{s}^{(0)}$ is the charge carrier density for $U=U\sub{CPD}$ and $n\sub{s}(U)$ is the voltage dependent charge carrier density in the 2DEG underneath the tip.

For total depletion under the tip $n\sub{s}(U\sub{depl})=0$ and the depletion voltage is
\begin{equation}
  U\sub{depl} = U\sub{CPD} + \frac{e\,n\sub{s}^{(0)}}{\epsilon_{0}\epsilon_{1}\epsilon_{2}}\left (
     \epsilon_{1}D + \epsilon_{2}z  \right),
   \label{eq:ns}
\end{equation}
i.e. there is a linear dependence between depletion voltage and tip-sample
separation $\Delta z$. The free parameter determining the slope of $U\sub{depl}(z)$ is the electron
density $n\sub{s}^{(0)}$ of the 2DEG.

From the data plotted in \fref{fig:delta-mu}a, a local electron
density of $n\sub{s}\ttop{local} = 1.9\cdot10^{15}\,{\rm m}^{-2}$ is
extracted from the slope of the data points.  This compares to the
electron densities gained from Shubnikov-de Haas and Hall transport
measurements. They are $\nHall = 1.70\times 10^{-15}{\rm m}^{-2}$ and
$\nSdH = 1.56\times 10^{-15}{\rm m}^{-2}$. The corresponding curves in
\fref{fig:delta-mu}a have been generated using equation \eqref{eq:ns}. 

The three results differ slightly. Considering that the methods and
scopes of the three measurements are different, this is not
unexpected.  The local measurement probes the local properties of
the electron gas right underneath the tip whereas transport
measurements average over the whole sample area.  Scanning electron
microscope images of the tip performed after warming suggest a tip
radius $R$ in the range of 1$\,\mu$m. This is more than an order
of magnitude larger than the average tip-sample separation and hence
the plate capacitor model is justified.

When talking about local measurements, the question of the lateral
resolution arises. We have not yet performed scanning capacitance
experiments with the described method, but as the method relies on
$R\gg\Delta z$ the resolution will be limited by the tip-radius $R$.

%
The contact potential difference $U\sub{CPD}$ between the PtIr-tip
and the 2DEG can be extracted from positions of the maxima of the
fitted parabolae. In \fref{fig:delta-mu} we plot $U\sub{CPD}$ versus
the tip-sample separation.  There is a slight decrease of
$U\sub{CPD}$ with $\Delta z$. The typical value for $U\sub{CPD}$ for
a PtIr-heterostructure system is $0.5\,$V as measured. This value is
important because it has to be taken into account in non-invasive
electronic measurements.

Only the high stiffness of a tuning fork oscillator allows for the
presented measurements. Although softer cantilevers suggest a higher
force resolution, they bend with attracting forces and $\Delta z$
would no longer be constant. At higher forces the tip on a soft
cantilever would even stick to the sample in what is generally known
as ``snap-in''.

A general model not reproduced here involving doping ions and
surface charges adds a distance dependence to the expression for
$U\sub{CPD}$.  Reducing it to a plate capacitor cancels out this
dependence.


In conclusion, we have performed low-temperature local Kelvin probe
measurements on a Al[Ga]As heterostructure using a tuning fork based
scanning probe microscope.  With the help of a plate capacitor model
the local electron density underneath the tip could be determined. 

The authors acknowledge financial support from the Swiss National
Science Foundation.


\end{document}